\documentclass[aps,prd,twocolumn,nofootinbib,showpacs]{revtex4}

\usepackage[dvips]{graphicx} \usepackage{mathbbol} \usepackage[sort&compress]{natbib}
\begin{document}
\title{Non-thermal leptogenesis with strongly hierarchical right handed neutrinos}
\author{V. Nefer \c{S}eno\u{g}uz} \email{nefer@ku.edu}
\affiliation{Department of Physics and Astronomy, University of Kansas, Lawrence, KS 66045, USA} 
\begin{abstract}
Assuming the Dirac-type neutrino masses $m_D$ are related to quark or charged
lepton masses, neutrino oscillation data indicate that  right handed neutrino
masses are in general strongly hierarchical. In particular, if $m_D$ is similar
to the up-type quark masses,  the mass of the lightest right handed neutrino
$M_1\lesssim10^6$ GeV.  We show that non-thermal leptogenesis by inflaton decay
can yield sufficient baryon asymmetry despite this constraint, and discuss how
the asymmetry is correlated with the low energy neutrino masses and CP-violating
phases.
\end{abstract}
\maketitle

\section{Introduction}
An attractive mechanism for generating the observed baryon asymmetry of the
universe (BAU) is baryogenesis via leptogenesis \cite{Fukugita:1986hr}. In the
seesaw model \cite{Minkowski:1977sc},
the out-of-equilibrium decays of right handed (RH) neutrinos to lepton and
Higgs fields create lepton asymmetry, which is partially converted to baryon
asymmetry by electroweak sphaleron processes \cite{Kuzmin:1985mm}. 

The RH neutrinos can be generated thermally after inflation,
if their masses are comparable to or below the reheat temperature $T_r$.
The thermal leptogenesis scenario has the nice feature that the final asymmetry is independent
of initial conditions and inflaton couplings. However, it requires $T_r\gtrsim10^9$ GeV
to generate the BAU \cite{Giudice:2003jh,Buchmuller:2004nz}, which is problematic in supersymmetric (SUSY) models
due to the gravitino constraint \cite{Khlopov:1984pf}.
Non-thermal leptogenesis by inflaton decay is an alternative scenario that can work
with lower values of $T_r$ ($\gtrsim10^6$ GeV) \cite{Lazarides:1991wu,Giudice:1999fb,Asaka:1999yd}.
These bounds can be saturated with $M_1\sim T_r$ and $M_1\gtrsim T_r$ for the thermal and non-thermal
scenarios respectively, where $M_1$ is the lightest RH neutrino mass.

The seesaw relation
\begin{equation} \label{ssw}
m=m_D M^{-1} m^T_D\,,
\end{equation}
where $m_D$ is the Dirac-type neutrino mass matrix, relates the RH neutrino mass matrix $M$
to the low energy neutrino mass matrix $m$, given in the basis where the charged lepton
mass matrix and gauge interactions are diagonal by
\begin{equation} \label{pmns}
m=U^*_{\rm PMNS}d_{\nu} U^{\dag}_{\rm PMNS}\,.
\end{equation}
Here $d_{\nu}\equiv diag(m_1,m_2,m_3)$, and $U_{\rm PMNS}$ \cite{Pontecorvo:1957cp} 
is the leptonic mixing matrix
\begin{equation}
{\setlength\arraycolsep{2pt}
{\scriptsize
\left( \begin{array}{ccc}
 c_{12}c_{13} & s_{12}c_{13} & s_{13} e^{-i\delta} \\ 
-c_{23}s_{12}-s_{23}c_{12}s_{13} e^{i\delta} &
 c_{23}c_{12}-s_{23}s_{12}s_{13} e^{i\delta} & s_{23}c_{13} \\ 
 s_{23}s_{12}-c_{23}c_{12}s_{13} e^{i\delta} & 
-s_{23}c_{12}-c_{23}s_{12}s_{13} e^{i\delta} & c_{23}c_{13}
\end{array} \right)}}\!\cdot K_0\,,
\end{equation}
$c_{ij} \equiv \cos \theta_{ij}$, $s_{ij} \equiv \sin \theta_{ij}$,
$\delta$ is the CP-violating Dirac phase and 
$K_0=diag(e^{i\alpha_1/2},e^{i\alpha_2/2},1)$ 
contains the two CP-violating Majorana
phases. 

In Refs. \onlinecite{Branco:2002kt,Berger:1999bg,Akhmedov:2003dg,Vives:2005ra,Hosteins:2006ja}, 
thermal leptogenesis was analyzed with the assumption that $m_D$ is related to
the mass matrices of quarks or charged leptons, as typically realized in grand
unified theories. In this case the Dirac masses are hierarchical, and the Dirac
left-handed rotation in the basis where the charged lepton mass matrix is diagonal (the leptonic
analogue of $U_{\rm CKM}$) is expected to be nearly diagonal or similar to $U_{\rm CKM}$.
We will hereafter refer to these two assumptions as quark-lepton symmetry.

Hierarchical Dirac masses indicate
strongly hierarchical RH neutrino masses \cite{Smirnov:1993af,Akhmedov:2003dg},
and the resulting BAU is suppressed due to the low value of $M_1$.
In particular, $M_1\lesssim10^6$ GeV if $m_D$ is similar to the up-type quark masses. 
In this letter we point out that sufficient asymmetry can nevertheless be generated
through non-thermal leptogenesis by inflaton decay. The inflaton is assumed
to decay predominantly to the next-to-lightest RH neutrino. The asymmetry resulting
from decays of this neutrino is partially washed out since $M_1< T_r$. The
final asymmetry depends on the asymmetry per neutrino decay as well as how
strong the washout is.

The plan of the paper is as follows: We first review
the structure of seesaw parameters and estimate the asymmetry and the washout
assuming quark-lepton symmetry. Numerical examples are provided in separate sections
for normal and inverted hierarchical (or quasi-degenerate) light neutrino masses. We discuss how the BAU
is correlated with the CP-violating phases and conclude
with a summary of results and some brief remarks on thermal leptogenesis.

\section{Seesaw parameters and leptogenesis}
In the basis where the RH neutrino mass matrix is diagonal, the Dirac
mass matrix can be written as
\begin{equation} \label{mdd}
m_D=U^{\dag}_L d_{D} U_R\,,
\end{equation}
$d_D\equiv diag(m_{D1},m_{D2},m_{D3})$. Eq. (\ref{ssw}) then takes the form
\begin{equation} \label{uwu}
m=U^{\dag}_Ld_{D}Wd_{D}U^*_L\,,
\end{equation}
where
\begin{equation} \label{rwr}
W\equiv U_R d^{-1}_R U^T_R
\end{equation}
is the inverse mass matrix of the RH neutrinos 
in the basis where $m_D=U^{\dag}_Ld_{D}$, and $d_R\equiv diag(M_1,M_2,M_3)$. From
Eq. (\ref{uwu}) one obtains 
\begin{equation}
W= 
\left(\begin{array}{ccc}
\vspace*{0.2cm}
\frac{\hat{m}_{ee}}{m_{D1}^2} & \frac{\hat{m}_{e\mu}}{m_{D1}m_{D2}} & \frac{\hat{m}_{e\tau}}{m_{D1}m_{D3}}\\
\vspace*{0.2cm}
\dots & \frac{\hat{m}_{\mu\mu}}{m_{D2}^2} & \frac{\hat{m}_{\mu\tau}}{m_{D2}m_{D3}}\\
\vspace*{0.2cm}
\dots & \dots & \frac{\hat{m}_{\tau\tau}}{m_{D3}^2} \end{array}\right) \,,
\end{equation}
\begin{equation} \label{mrhat}
\hat{m}\equiv U_L m U_L^T\,.
\end{equation}
As mentioned in the introduction, we are assuming $m_{D1}\ll m_{D2}\ll m_{D3}$, and
the Dirac left-handed rotation $U_L\approx U_{\rm CKM}\approx\mathbb{1}$.
Elements of $\hat{m}\approx m$ generally have a much milder hierarchy 
compared to the Dirac masses. The matrix $W$
then has a simple hierarchical structure, and is diagonalized by \cite{Akhmedov:2003dg}
\begin{equation} \label{urr}
U_R \approx{\small \left(
\begin{array}{ccc}
\vspace*{0.2cm}
1 & -\left(\frac{\hat{m}_{e\mu}}{\hat{m}_{ee}}\right)^*\frac{m_{D1}}{m_{D2}} &
\left(\frac{d_{23}}{d_{12}}\right)^*\frac{m_{D1}}{m_{D3}} \\
\vspace*{0.2cm}
\left(\frac{\hat{m}_{e\mu}}{\hat{m}_{ee}}\right)\frac{m_{D1}}{m_{D2}} & 1 &
-\left(\frac{d_{13}}{d_{12}}\right)^*\frac{m_{D2}}{m_{D3}} \\
\vspace*{0.2cm}
\left(\frac{\hat{m}_{e\tau}}{\hat{m}_{ee}}\right)\frac{m_{D1}}{m_{D3}} &
\left(\frac{d_{13}}{d_{12}}\right)\frac{m_{D2}}{m_{D3}} & 1 
\end{array}\right)}\!\cdot K\,,
\end{equation}
\begin{eqnarray} \nonumber
d_{23}&\equiv& \hat{m}_{e\mu} \hat{m}_{\mu\tau}-\hat{m}_{\mu\mu}\hat{m}_{e\tau}\,,\\ \nonumber
d_{13}&\equiv& \hat{m}_{ee}\hat{m}_{\mu\tau}-\hat{m}_{e\mu}\hat{m}_{e\tau}\,,\\ \nonumber
d_{12}&\equiv& \hat{m}_{ee}\hat{m}_{\mu\mu}-\hat{m}_{e\mu}^2\,,\\ \nonumber
K&=&diag(e^{-i\phi_1/2},e^{-i\phi_2/2},e^{-i\phi_3/2})\,,\quad
\phi_i \equiv \arg{M_i} \,.
\end{eqnarray}
Here the phases of RH neutrinos $\phi_i$ are included in $U_R$ to keep $M_i$ real.
The mass eigenvalues are
\begin{equation} \label{rhm}
M_1\approx  \left|\frac{m_{D1}^2}{\hat{m}_{ee}}\right|\,,\
M_2\approx \left|\frac{m_{D2}^2 \hat{m}_{ee}}{d_{12}}\right|\,,\
M_3\approx \left|\frac{m_{D3}^2d_{12}} {m_1m_2m_3}\right|\,.
\end{equation}
The large neutrino mixings can originate from the seesaw, despite both $U_L$ 
and $U_R$ being nearly diagonal \cite{Smirnov:1993af}. 

To estimate the BAU, suppose the inflaton predominantly decays into the $i$-th family RH neutrino $N_i$.
The comoving number density $Y_{N}$ is given by
\begin{equation}
Y_{N}\equiv\frac{n_{N}}{s}=\frac{n_{N}}{n_{\phi}}\frac{n_{\phi}}{\rho_{\phi}}\frac{\rho_{\phi}}{s}
=2Br\frac{1}{m_{\phi}}\frac{3T_r}{4}\,,
\end{equation}
$Br\le1$ is the branching ratio of the inflaton $\phi$ to $N_i$, the factor 2 assumes $\phi\to2N_i$, 
$m_{\phi}$ is the inflaton mass, and we have used the instantaneous decay approximation. A more accurate calculation
shows $Y_{N}$ to be $\approx25\%$ larger \cite{Kolb:1990vq}.
The asymmetry resulting from the decays of $N_i$ (assuming 
it decays promptly \cite{Giudice:1999fb}) is then
\begin{equation} \label{lepp}
Y_{\Delta}\lesssim\frac{2T_r|\epsilon_i|\eta}{m_{\phi}}\,,
\end{equation}
where $\Delta\equiv(1/3)B-L$, $\epsilon_i$ is the lepton asymmetry 
produced per decay of $N_i$, and  $\eta$ is a washout factor. In the simplest scenario, 
$M_1\gg T_r$ and there is no washout ($\eta=1$). 
On the other hand, if $M_1\lesssim T_r$, part of the asymmetry will be washed out due to
$N_1$ mediated inverse decays and $\Delta L=1$ scatterings. 
For $M_2\lesssim T_r$, $N_2$ mediated processes contribute
to the washout as well.
The $\Delta$ asymmetry 
is multiplied by a conversion factor ($C\approx12/37$ for SM and $C\approx10/31$ for MSSM)
to obtain the BAU resulting from sphaleron processes 
at equilibrium above the electroweak scale \cite{Khlebnikov:1988sr}.

For hierarchical RH neutrino masses as in Eq. (\ref{rhm}),
\begin{equation} \label{davidson}
|\epsilon_1|\le\frac{3aM_1m_{\rm atm}}{16\pi v^2}\,,
\end{equation}
where the parameter $a=1$ for non-SUSY and $a=2/\sin^2\beta$ for SUSY
($\tan\beta\equiv\langle H_u\rangle/\langle H_d\rangle$) and
$v=174$ GeV \cite{Asaka:1999yd,Hamaguchi:2001gw,DiBari:2005st}.
Eqs. (\ref{lepp}, \ref{davidson}) imply that if the inflaton decays into $N_1$,
the WMAP best fit $Y_{B0}=8.7\times10^{-11}$ \cite{Spergel:2006hy} for the BAU requires
\begin{equation} \label{reh2}
M_1\gtrsim\left(\frac1{a\eta}\right)\left(\frac{1/3}{C}\right)\left(\frac{0.05{\rm\ eV}}{m_{\rm atm}}\right)
1.3\times10^6{\rm\ GeV}\,,
\end{equation}
since $m_{\phi}\gtrsim T_r$ (in effect \cite{Kolb:2003ke}).

If the Dirac masses are related to the up-type quark masses, Eq. (\ref{rhm}) indicates
that $M_1$ is too light to generate the BAU. We will therefore assume $m_{\phi}>2M_2$
so that $\phi$ predominantly decays into $N_2$ instead of $N_1$.
Using Eqs. (\ref{mdd}, \ref{urr}), $(m_D^{\dag}m_D)_{ij}\sim m_{Di}m_{Dj}$ 
with coefficients involving elements of $\hat{m}$. It follows that for
hierarchical Dirac masses the dominant contribution to the asymmetry from the decays of $N_2$ involves
$N_3$ in the loop  \cite{Senoguz:2003hc}:
\begin{eqnarray} \label{ep1} \epsilon_2&\equiv&\sum_{\alpha}\epsilon_{2,\alpha}\\ \nonumber &\approx&-
   \frac{3a}{16\pi v^2} \frac{\sum_{\alpha} {\rm Im} \left[ (m_D^{\dag})_{2\alpha} (m_D^{\dag}m_D)_{23}(m^T_D)_{3\alpha} \right]}
{\left(m_D^{\dag} m_D\right)_{22}}    
\frac{M_2}{M_3} \,.  
\end{eqnarray}
In the expression $\epsilon_{2,\alpha}$, the label $i=1,2,3$ refers to the RH
neutrino, and $\alpha=e,\mu,\tau$ to the lepton flavor that it decays into.
For quark-lepton symmetry (that is, also assuming $U_L\approx\mathbb{1}$),
the Dirac mass matrix has the  form
\begin{equation} \label{mdir}
m_D\sim\left(\begin{array}{ccc}
\vspace*{0.2cm}
{\cal O}(m_{D1}) & \ll m_{D2} &\ll m_{D3} \\
\vspace*{0.2cm}
{\cal O}(m_{D1})&{\cal O}(m_{D2}) & \ll m_{D3}\\
\vspace*{0.2cm}
{\cal O}(m_{D1})&{\cal O}(m_{D2})  &{\cal O}(m_{D3})  \\ \end{array}\right) \,,
\end{equation}
with coefficients involving elements of $\hat{m}$,
and the terms above the main diagonal proportional to non-diagonal
elements of $U_L$. 
It follows that the dominant term in Eq. (\ref{ep1}) is
\begin{eqnarray} \label{eptau}|\epsilon_{2,\tau}|&\approx&
   \frac{3a\varphi}{16\pi v^2} \frac{\left|m_{D32}^{*} m_{D33} \right|^2}
{\left(m_D^{\dag} m_D\right)_{22}}    
\frac{M_2}{M_3} \,,   \\ \label{eptau2}
&\approx&\frac{3a\varphi M_2}{16\pi v^2}\frac{|d_{13}|^2 m_1 m_2 m_3}{|d_{12}|(|d_{12}|^2+|d_{13}|^2)}\,,
\end{eqnarray}
where $\varphi\le1$ is an effective phase that depends on 
$d_{\nu},\ U_{\rm PMNS},\ d_D$ and $U_L$.
(The phases $\phi_i$ in $U_R$ can be calculated using Eqs. (\ref{uwu}, \ref{rwr})
given the above masses, mixings and phases.)

To estimate $\eta_i$ (the washout involving $N_i$), we define the washout parameters
\begin{equation} \label{washiyo}
K_{i,\alpha}\equiv\frac{\tilde{m}_{i,\alpha}}{m_*},\quad\tilde{m}_{i,\alpha}\equiv\frac{|m_{D\alpha i}|^2}{M_i}\,,
\end{equation}
$m_*\approx1.08\times10^{-3}$ eV for  non-SUSY and $m_*\approx(\sin^2\beta)1.58\times10^{-3}$ eV
for  SUSY. Note that  lepton flavors should be treated separately for an accurate calculation of the
washout \cite{Barbieri:1999ma,Vives:2005ra,Abada:2006fw,Abada:2006ea}.
Setting $\alpha=\tau$, the washout is given in the instantaneous decay approximation by \cite{Buchmuller:2004nz,Abada:2006ea}
\begin{equation} \label{washh}
\eta_{i,\tau}\approx\exp\left[\int_{z_0}^{\infty}-\frac14z^3\mathcal{K}_1(z)j(z)K_{i,\tau}A_{\tau\tau}{\rm d}z\right]\,,
\end{equation}
$z\equiv M_i/T$, $z_0\equiv M_i/T_r$, $\mathcal{K}_1$ is a modified Bessel function of the second kind, and
\begin{equation}
Y_{\ell\alpha}=-A_{\alpha\beta}Y_{\Delta\beta}\,,
\end{equation}
with $\ell$ denoting the lepton doublet. 
The value of $A_{\tau\tau}$ depends on
which interactions are in thermal equilibrium \cite{Barbieri:1999ma}.
For MSSM, $A_{\tau\tau}=19/30$ between $(1+\tan^2\beta)\times10^5$ GeV
and $(1+\tan^2\beta)\times10^9$ GeV \cite{Antusch:2006cw}. For SM, $A_{\tau\tau}=344/537$
and $390/589$ below and above $10^9$ GeV respectively \cite{Abada:2006ea}.\footnote{A 
more accurate analysis around this temperature should take  
quantum oscillations into account \cite{Abada:2006fw}.}

The function $j(z)$ takes $\Delta L=1$ scatterings  into account. We will not attempt
a detailed calculation which involves finite temperature effects. Instead, we will
use $j(z)=1$ to define $\eta_{\rm max}$ which underestimates the washout, and
\begin{equation} \label{jz}
j(z)=\frac{\mathcal{K}_2(z)}{\mathcal{K}_1(z)}\left(\frac{9m_t^2}{8\pi^2v^2z}+1\right)
\end{equation}
to define $\eta_{\rm min}$ which overestimates the washout \cite{Buchmuller:2004nz}.

It is also required that $\Delta L=2$ processes mediated by RH neutrinos are out of
equilibrium. As discussed in Ref. \onlinecite{Giudice:1999fb}, it is sufficient to have 
$T_r\lesssim(m_{\rm atm}/m_i)^210^{13.5}$ GeV provided
\begin{equation} \label{gam}
\frac{\Gamma_{N_2}}{\Gamma_{\phi}}=\frac{(a/8\pi v^2)\sum_{\alpha}|m_{D\alpha2}|^2M_2}
{\sqrt{2\pi^2g_*/45}T_r^2/m_P}>1\,,
\end{equation}
where $m_P\approx2.4\times10^{18}$ GeV is the reduced Planck scale, and the relativistic degrees of freedom 
$g_*=106.75$ (228.75) for SM (MSSM).
Using $\sum_{\alpha}\tilde{m}_{2,\alpha}\sim m_{\rm atm}$, this condition
corresponds to $M_2\gtrsim T_r/5$.

For $M_2\gtrsim T_r$, $\Gamma_{N_2}\gg\Gamma_{\phi}$ and we can use the following simplified
equations \cite{Giudice:2003jh,Antusch:2006cw,Antusch:2006gy}:
\begin{eqnarray} \label{z}
Z\frac{{\rm d}\rho_{\phi}}{{\rm d}z}&=&-\frac3z\rho_{\phi}-\frac{\Gamma_{\phi}}{Hz}\rho_{\phi} \,, \\ \label{zx}
ZX\frac{{\rm d}Y_{\Delta_{\tau}}}{{\rm d}z}&=&\frac3z(Z-1)XY_{\Delta_{\tau}}
+\frac{2\Gamma_{\phi}\rho_{\phi}X\epsilon_{2,\tau}}{sHzm_{\phi}} \\ \nonumber
&-&\frac14z^3{\cal K}_1(z)j(z)K_{2,\tau}A_{\tau\tau}Y_{\Delta_{\tau}} \\ \nonumber
&-&\frac14\gamma (\gamma z)^3{\cal K}_1(\gamma z)j(\gamma z)K_{1,\tau}A_{\tau\tau}Y_{\Delta_{\tau}} \,.
\end{eqnarray}
Here $z\equiv M_2/T$, $\gamma\equiv M_1/M_2$, and
\begin{equation} \label{zazz}
Z\equiv1-\frac{\Gamma_{\phi}\rho_{\phi}}{4H\rho_r}\,,\quad X\equiv\left(\frac{\rho_r+\rho_{\phi}}{\rho_r}\right)^{1/2}\,,
\end{equation}
with $\rho_r=(M_2/z)^4g_*\pi^2/30$. The equations are solved from $z_i=M_2/T_{\rm max}$ to $z_f\gg\gamma^{-1}$, and
$Y_B\approx CY_{\Delta_{\tau}}(z_f)$.\footnote{$T_{\rm max}
\sim(H_I m_P)^{1/4}T_r^{1/2}$ is the maximum temperature attained just after the inflaton starts oscillating,
at time $H_I^{-1}$. We took $H_I=m_{\phi}$ for the numerical calculation, but the results are not sensitive
to $H_I$ as long as $T_{\rm max}$ is at least a few times larger than $T_r$ ($H_I\gg T_r^2/m_P$).}

\section{Results for NH spectrum}
In this section we assume a normal hierarchical (NH) spectrum of
light neutrino masses ($m_3\approx m_{\rm atm}, m_2\approx m_{\odot}$, $m_1\ll m_2$).
To simplify the discussion we also set $U_L=\mathbb{1}$ and $s_{13}=0$.
In this limit the RH neutrino masses are given by \cite{Branco:2002kt,Akhmedov:2003dg}
\begin{equation} \label{masses}
M_1\approx\frac{m^2_{D1}}{s^2_{12}m_2}\,,\quad M_2\approx\frac{2m_{D2}^2}{m_3}\,,\quad M_3\approx\frac{m_{D3}^2s^2_{12}}{2m_1}\,,
\end{equation}
and with $|d_{12}|=|d_{13}|=s^2_{12}m_2m_3/2$ we obtain
\begin{equation} \label{eps2}
|\epsilon_{2,\tau}|\approx\frac{3a\varphi m_1M_2}{16\pi v^2 s^2_{12}}\,.
\end{equation}
Using Eqs. (\ref{lepp}, \ref{eps2}),
\begin{eqnarray} \label{madmax}
Y_B&\approx&\left(\frac{2M_2}{m_{\phi}}\right)\left(\frac{m_1}{m_2}\right)\varphi Y_{B,{\rm max}}\,,\\
\label{barbar}
Y_{B,{\rm max}}&\approx&\frac{3aCT_rm_2\eta_1\eta_2}{16\pi v^2 s^2_{12}}\,.
\end{eqnarray}
For $M_1\ll T_r$ we can take $z_0=0$ in Eq. (\ref{washh}) to obtain
\begin{eqnarray} \label{wasminmax}
\eta_{\rm 1,max}&=&\exp\left[-\frac{3\pi}{8}K_{1,\tau}A_{\tau\tau}\right]\,,\\ \nonumber
\eta_{\rm 1,min}&=&\exp\left[-\left(2+\frac{3\pi}{8}\frac{9m_t^2}{8\pi^2v^2}\right)K_{1,\tau}A_{\tau\tau}\right]\,.
\end{eqnarray}
$\eta_2$ can be estimated by using
Eq. (\ref{washh}), and becomes significant for
$z_0\lesssim10$.  It follows from Eqs. (\ref{madmax}, \ref{barbar}) that the maximum
asymmetry is obtained for $T_r\approx M_2/10$ and
$m_{\phi}\approx20T_r$.  Taking the reheating phase into account by solving
the Boltzmann equations gives similar results.

\begin{figure}[t]
\includegraphics[height=.225\textheight]{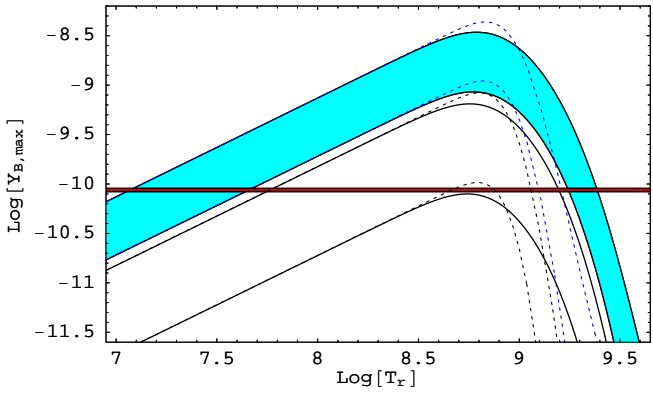}
\vspace{-.6cm}
\caption{$Y_{B,{\rm max}}$ versus the reheat temperature $T_r$, for $d_D=d_u$, $U_L=\mathbb{1}$ and $s_{13}=0$. 
The horizontal band corresponds to the WMAP range $(8.7\pm0.3)\times10^{-11}$ \cite{Spergel:2006hy}, 
the solid and the dotted curves are calculated using Eqs. (\ref{z}--\ref{zazz}) and Eq. (\ref{washh})
respectively. Filled: SUSY, unfilled: non-SUSY; upper bounds: $j(z)=1$, lower bounds:
Eq. (\ref{jz}).
} \label{fig}
\end{figure}

A numerical example is shown in Fig. \ref{fig}, where we have used Eq.
(\ref{barbar}) and set $d_{D}=d_{u}\equiv diag(m_u,m_c,m_t)$ with the values
$m_u=1.5$ MeV, $m_c=0.43$ GeV, $m_t=150$ GeV (taken from Ref.
\onlinecite{Fusaoka:1998vc}, for  a renormalization scale of $10^9$ GeV), for which
$M_1\approx6\times10^5$ GeV and $M_2\approx6\times10^9$ GeV.\footnote{We take
$s_{12}=1/\sqrt{3}$, $s_{23}=1/\sqrt{2}$  and $\sin\beta\approx1$ in the numerical calculations. We
also take  $m_3=0.06$ eV and $m_2=0.011$ eV, roughly approximating
renormalization group effects by increasing the neutrino mass scale 20\%.}
Assuming $m_1\ll m_2$, we can ignore the contributions to
$\tilde{m}_{i,\alpha}$ that involve $m_1$, and it follows from Eq.
(\ref{madmax}) that $Y_B\propto m_1$.  

While the washout due to $N_2$ is  severe when $M_2\lesssim T_r$, the washout
due to $N_1$ is rather mild, of order 0.1, for $U_L=\mathbb{1}$ and
$s_{13}=0$. This follows from Eq. (\ref{washh}), with
$\tilde{m}_{2,\tau}\approx m_3/2$ compared to
$\tilde{m}_{1,\tau}\approx c_{12}^2m_2/2$ (in the limit $m_1\ll
m_{2,3}$).  For $U_L\approx U_{\rm
CKM}$, there are additional, order $\theta_C^2 m_3^2/m_2$ contributions to
$\tilde{m}_{1,\tau}$, where $\theta_C$ is the leptonic analogue of the Cabibbo angle. The result then
depends on the CP-violating phases of $U_{\rm PMNS}$, but on average the washout
gets stronger.

We also take into account an effect due to off-diagonal elements of $A_{\alpha\beta}$ \cite{Shindou:2007se}.
Namely, in case of a strong washout related to a large $\tilde{m}_{1,\tau}$, part of the asymmetry
can still survive if $\tilde{m}_{1,\mu}$ or $\tilde{m}_{1,e}\sim m_*$.\footnote{This is also true for
the washout due to $N_2$, and part of the asymmetry can survive for $M_2\lesssim T_r$.
However, the maximum BAU is still obtained for $M_2\approx10T_r$.} 
Typically $\tilde{m}_{1,e}$ is the smallest washout parameter.
For an estimate we can ignore $\tilde{m}_{1,\mu}$ which is  of order $\tilde{m}_{1,\tau}$, and modify the Boltzmann
equations by adding
\begin{equation}
-\frac14\gamma (\gamma z)^3{\cal K}_1(\gamma z)j(\gamma z)K_{1,\tau}A_{\tau e}Y_{\Delta_{e}}
\end{equation}
to Eq. (\ref{zx}), and including an analogous equation for $Y_{\Delta_e}$
(with $\tau\leftrightarrow e$ and $\epsilon_{2,e}\approx0$). The final asymmetry is
then $Y_B\approx C(Y_{\Delta_{\tau}}(z_f)+Y_{\Delta_{e}}(z_f))$.

\begin{figure}[t]
\includegraphics[height=.225\textheight]{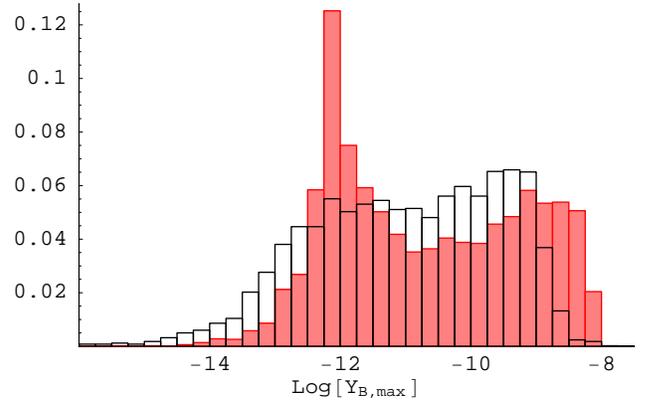}
\vspace{-.6cm}
\caption{Histograms for $Y_{B,{\rm max}}$ with $d_D=d_u$, $m_1=0.2m_2$ and $U_L=U_{\rm CKM}$, calculated for
 SUSY  with $j(z)=1$. Filled: $s_{13}=0$, unfilled: $s_{13}=0.2$.
} \label{fig2}
\end{figure}

To estimate the probability distribution of $Y_{B,{\rm max}}$ for $U_L=U_{\rm CKM}$, we 
numerically solved the Boltzmann equations 5000 times with uniformly distributed random phases of $U_{\rm PMNS}$.
We define $Y_{B,{\rm max}}$ by taking $m_{\phi}=2M_2$ as in Eq. (\ref{madmax}), but here we take $m_1=0.2m_2$
to be specific and include $\varphi$. 
In addition, the asymmetry is maximized by varying $T_r$ for each run.
Fig. \ref{fig2} shows the results for $s_{13}=0$ and $s_{13}=0.2$.
The percentage of runs yielding $Y_{B,{\rm max}}>Y_{B0}$
was 38\% and 32\% for $s_{13}=0$ and $s_{13}=0.2$
respectively. Including the $A_{\tau e}$ term significantly alters the low end of the
probability distribution for $Y_{B,{\rm max}}$, but the effect on these percentages is only a
few points.\footnote{Note that in Eq. (\ref{mdd}), there are generally Majorana phases on the left side
of $U_R$ as well. These phases enter $\tilde{m}_{i,\alpha}$, 
leading to ${\cal O}(\theta_C)$ corrections for $\alpha=e,\mu$. 
(They can be rotated away in the limit $U_L=\mathbb{1}$.) This
does not affect the results appreciably.}

\begin{figure}[t]
\includegraphics[height=.225\textheight]{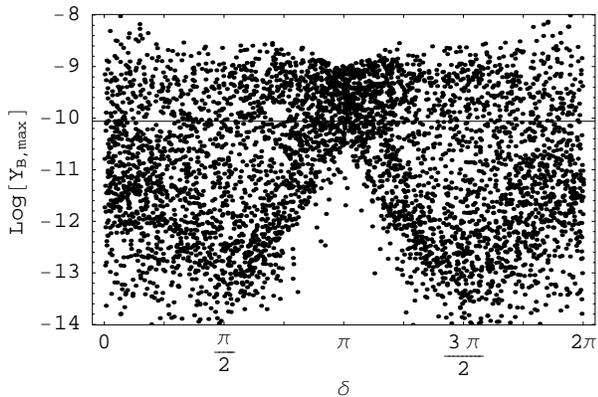}
\vspace{-.6cm}
\caption{$Y_{B,{\rm max}}$ versus the Dirac phase $\delta$, with $d_{D}=d_{u}$, $m_1=0.2m_2$, $U_L=U_{\rm CKM}$ and  
$s_{13}=0.2$, calculated for SUSY  with $j(z)=1$.
} \label{fig3}
\end{figure}

For $s_{13}=0$, the peak at $Y_B\approx10^{-12}$
results from $\tilde{m}_{1,e}$ having a relatively small deviation
($\tilde{m}_{1,e}\approx[1+{\cal O}(\theta_C)+{\cal O}(\theta^2_Cm_3/m_2)]s^2_{12}m_2$).
For $s_{13}\ne0$, there are additional contributions to $\tilde{m}_{1,e}$
(as well as $\tilde{m}_{1,\tau}$), and the probability distribution becomes more dispersed. 
Using Eqs. (\ref{washiyo}, \ref{mdd}, \ref{urr}, \ref{rhm}),
$\tilde{m}_{1,\tau}\approx|\hat{m}_{e\tau}|^2/|\hat{m}_{ee}|$, which from Eqs.
(\ref{mrhat}, \ref{pmns}) is  given by
\begin{equation}
\approx\frac{\left|-c_{12}s_{12}e^{i\alpha_2}m_2 
+\left(s_{13}e^{-i\delta}+\frac{\theta_C}{\sqrt{2}}\right)m_3\right|^2}
{2\left|s^2_{12}e^{i\alpha_2}m_2+\left(s_{13}e^{-i\delta}+\frac{\theta_C}{\sqrt{2}}\right)^2m_3\right|}\,.
\end{equation}
The terms including $m_1$ and $\alpha_1$ are subdominant. 
Assuming $\theta_C$ is similar to the Cabibbo angle, 
$\tilde{m}_{1,\tau}$ is minimized for  $\delta\approx\pi$.
$Y_{B,{\rm max}}$ also depends on $\epsilon_{2,\tau}$,
which from Eq. (\ref{eptau2}) is maximized for $\delta\approx0$. For random Majorana phases,
$Y_{B,{\rm max}}>Y_{B0}$ is most likely at $\delta\approx\pi$ due to the exponential dependence on $\tilde{m}_{1,\tau}$,
however it remains possible for all values of $\delta$ (Fig. \ref{fig3}).

The values of $s_{13}$ and $\delta$ will be probed by neutrino beam experiments
within a decade for $s_{13}\gtrsim0.05$ \cite{Huber:2004ug}. The value
of $\alpha_2$ can in principle be probed by neutrinoless double beta decay
experiments. However, for normal hierarchy the effective Majorana mass 
$|\langle m_{\beta\beta}\rangle|=|m_{ee}|\approx|s^2_{12}e^{i\alpha_2}m_2+s^2_{13}e^{-2i\delta}m_3|$
is too small to detect using current techniques.

\section{Results for IH and QD spectra}
For inverted hierarchical (IH) spectrum of neutrino masses, $m_3\ll m_1<m_2\approx m_{\rm atm}$ and
in the limit $m_3\to0$, $U_L\to\mathbb{1}$,
\begin{equation}
\tilde{m}_{1,\tau}=\frac{c_{12}^2s_{12}^2\left|e^{i\alpha_1}m_1-e^{i\alpha_2}m_2\right|^2}
{2\left|c_{12}^2e^{i\alpha_1}m_1+s_{12}^2e^{i\alpha_2}m_2\right|}\,.
\end{equation}
Taking $s_{12}=1/\sqrt{3}$, $\tilde{m}_{1,\tau}$ ranges from $4m_{\rm atm}/3$ for $\alpha_2=
\alpha_1+\pi$ to $m_{\odot}^4/36m^3_{\rm atm}\approx0$ for $\alpha_2=
\alpha_1$. (Including the Cabibbo mixing, $\tilde{m}_{1,\tau}$ for $\alpha_2=
\alpha_1$ becomes $m_{\rm atm}\theta_C^2/4$, which is still $\lesssim m_*$.) 
As a result, the asymmetry is suppressed by a factor $\sim10^8$ for $\alpha_2=
\alpha_1+\pi$, but $Y_{B,{\rm max}}>Y_{B0}$ is 
possible if $\alpha_1\approx\alpha_2$.

The RH neutrino masses are given in this limit by
\begin{equation} \label{masses2}
M_1\approx\frac{m^2_{D1}}{m_{\rm atm}}\,,\quad M_2\approx\frac{2m_{D2}^2}{m_{\rm atm}}\,,
\quad M_3\approx\frac{m_{D3}^2}{2m_3}
\end{equation}
for $\alpha_1\approx\alpha_2$. With $|d_{12}|=|d_{13}|=m_1m_2/2$ we obtain
\begin{equation} \label{eps2b}
|\epsilon_{2,\tau}|\approx\frac{3a\varphi m_3M_2}{16\pi v^2 }\,,
\end{equation}
similar to Eq. (\ref{eps2}).

The asymmetry can only survive if $\alpha_1\approx\alpha_2$
for quasi-degenerate (QD) spectra of neutrino masses as well,
since terms involving $m_3$ in $\tilde{m}_{1,\tau}$ are suppressed either by $s_{13}$
or $\theta_C$. The RH neutrino masses $M_i\sim m_{Di}^2/\bar{m}$ where
$\bar{m}$ is the QD neutrino mass scale. Assuming $\alpha_1\approx\alpha_2$, 
$|\epsilon_{2,\tau}|\sim(3a/16\pi v^2)m_{D2}^2$ is maximized for $\alpha_2\approx\pi$.
On the other hand, $\tilde{m}_{1,\tau}\approx\theta_C^2 \bar{m}$ for $\alpha_2\approx\pi$
and it is minimized for $\alpha_2\approx0$. 
The maximum asymmetry is determined by the interplay of these two factors.

\begin{figure}[t]
\includegraphics[height=.225\textheight]{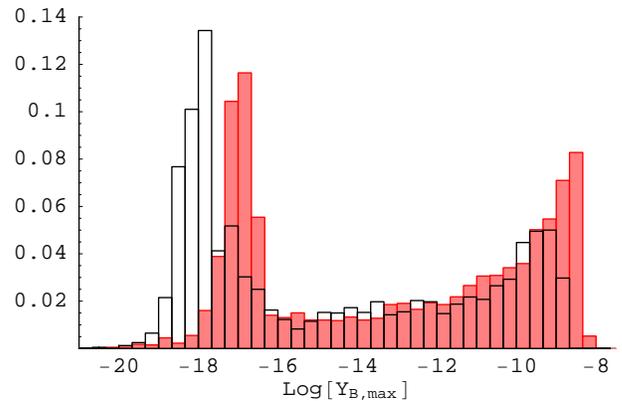}
\vspace{-.6cm}
\caption{Histograms for $Y_{B,{\rm max}}$ with $d_D=d_u$, $U_L=U_{\rm CKM}$ and $s_{13}=0$, calculated for
 SUSY  with $j(z)=1$. Filled: IH spectrum, unfilled: QD spectrum.
} \label{fig4}
\end{figure}

In the numerical examples we used the following neutrino masses. IH spectrum:
$m_1=0.059$ eV, $m_2=0.06$ eV, $m_3=m_2/5$. QD spectrum: $m_1=0.1$ eV, 
$m_2=0.1006$ eV, $m_3=0.117$ eV. (Similar results are obtained for inverted hierarchical QD masses.)
The resulting probability distribution of $Y_{B,{\rm max}}$
is displayed in Fig. \ref{fig4}. The percentage of runs yielding $Y_{B,{\rm max}}>Y_{B0}$
was 31\% and 18\% for IH and QD spectra respectively, for $U_L=U_{\rm CKM}$ and $s_{13}=0$. 
Since $\tilde{m}_{1,\tau}\propto\bar{m}$, $Y_{B,{\rm max}}$ decreases as $\bar{m}$ is increased,
with the  percentage of runs yielding  $Y_{B,{\rm max}}>Y_{B0}$ decreasing to 7\% (3\%) for $m_1=$\,0.2 (0.3) eV. 
These percentages increase a few points if $U_L\approx\mathbb{1}$, and decrease
a few points if $s_{13}\approx0.2$.

\begin{figure}[t]
\includegraphics[height=.225\textheight]{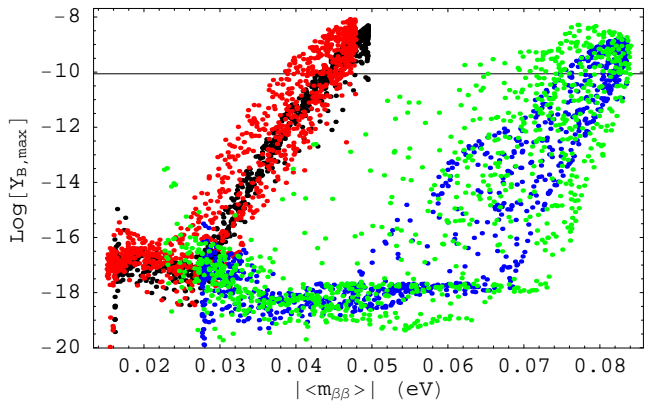}
\vspace{-.6cm}
\caption{$Y_{B,{\rm max}}$ versus the effective Majorana mass 
$|\langle m_{\beta\beta}\rangle|$, with $d_{D}=d_{u}$ and $U_L=U_{\rm CKM}$, calculated for SUSY  with $j(z)=1$.
Black (red): IH spectrum with $s_{13}=0$ ($s_{13}=0.2$). Blue (green): QD spectrum with
 $s_{13}=0$ ($s_{13}=0.2$). } \label{fig5}
\end{figure}

The effective Majorana mass, given by
\begin{equation}
|\langle m_{\beta\beta}\rangle|\approx\left|c_{12}^2e^{i\alpha_1}m_1+s_{12}^2e^{i\alpha_2}m_2\right|
\end{equation}
for both IH and QD spectra, is maximized by the condition $\alpha_2\approx\alpha_1$.
As shown in Fig. \ref{fig5}, $Y_{B,{\rm max}}>Y_{B0}$ requires $|\langle m_{\beta\beta}\rangle|\gtrsim0.04$ eV
for IH.\footnote{Note that since we took the neutrino
mass scale 20\% larger at the leptogenesis scale, we scaled $|\langle
m_{\beta\beta}\rangle|$ down 20\% in the figure to correspond to low energy
values.}
 This range of $|\langle m_{\beta\beta}\rangle|$ can be probed within a
decade \cite{Aalseth:2004hb}.

\section{Conclusion}
In this paper we considered non-thermal leptogenesis by inflaton
decay under the assumption that the Dirac-type neutrino mass matrix $m_D$ is related
to the up-type quark (or charged lepton) mass matrix. Following the approach of Ref. \onlinecite{Akhmedov:2003dg},
we did not make any specific assumptions on the textures of these matrices, but
rather considered the general structure that follows from fitting to the low
energy data. 
In this approach the RH neutrino masses are almost 
always strongly hierarchical ($\propto d_{D}^2$).

This strong hierarchy and the rest of our analysis follows from 
the light neutrino mass matrix having a less hierarchical structure compared to $d_D$,
except for specific values of $s_{13}$ and the $U_{\rm PMNS}$ phases
which occur very rarely in a random scan.\footnote{Since nothing is currently known about the values of 
these phases, we assumed a flat probability
distribution for numerical analysis. Such a distribution
also follows from `anarchy' \cite{Hall:1999sn}.} 
On the other hand, $m_D$ can be constrained further in particular $SO(10)$ or
other GUT models with flavor symmetries. It can then have structures different
from Eq. (\ref{mdir}), and fitting to the low energy data then
yields those specific values as predictions.  In such cases which are beyond
the scope of this paper, the RH neutrinos can be less
hierarchical and non-thermal leptogenesis with $M_1\gg T_r$ also becomes
possible.

Assuming strongly hierarchical RH neutrinos with $d_D\approx d_u$, the matter asymmetry created 
by the decays of $N_2$ is partially washed out 
since $M_1<T_r$, but can still account for the BAU. 
For either NH or IH spectra of light neutrino masses, 
$Y_{B}=Y_{B0}$ requires 
\begin{equation} \label{conc}
\left(\frac{10T_r}{M_2}\right)\left(\frac{2M_2}{m_{\phi}}\right)\left(\frac{{\rm min}(m_i)}{m_2}\right)\gtrsim
\left(\frac{m_c}{m_{D2}}\right)^2 10^{-2}\,,
\end{equation}
with each term on the left $<1$ (${\rm min}(m_i)=m_1$ for NH, ${\rm min}(m_i)=m_3$ for IH). 
The case $d_D=d_u$ corresponds to $M_2\sim6\times10^9$ GeV.\footnote{More precisely, $M_2$ varies depending on the
$U_{\rm PMNS}$ phases and is in the range $10^{9.7}$--$10^{10.2}$ GeV for NH with $s_{13}=0$.
For NH with $s_{13}=0.2$ or IH, it varies within an order (two orders) of magnitude for about 90\% (99\%) of the runs.}
Eq. (\ref{conc}) can then be satisfied with $m_{\phi}=10^{10}$--$10^{11}$ GeV
and $T_r\sim10^{8}$ GeV. This value of $T_r$ can be consistent with the gravitino constraint, while
$m_{\phi}$ in the above range is possible in  small field or hybrid inflation models. 
On the other hand, for simplest large field inflation models $m_{\phi}\gtrsim2\times10^{13}$ GeV.
The upcoming Planck satellite can discriminate these classes of models \cite{Dodelson:1997hr}.

For NH spectrum of light neutrino masses and $s_{13}\gtrsim0.1$, sufficient asymmetry is most likely
to be obtained if the $U_{\rm PMNS}$  Dirac phase $\delta\approx\pi$ (assuming $U_L\approx U_{\rm CKM}$).
For IH or QD spectra of light neutrino masses, sufficient asymmetry can only be
obtained if the $U_{\rm PMNS}$ Majorana phases are approximately equal to each other,
implying $|\langle m_{\beta\beta}\rangle|\approx m_{\rm atm}$ for IH spectrum and
larger for QD spectrum. The asymmetry decreases as the QD neutrino mass scale is
increased, and if $|\langle m_{\beta\beta}\rangle|\gtrsim0.2$ eV, the leptogenesis
scenario discussed here is strongly disfavored assuming $d_D\approx d_u$.
 
If we relate the Dirac masses to masses of the charged leptons $d_{\ell}\equiv diag(m_e,m_{\mu},m_{\tau})$ instead,
$m_{D2}\approx m_{\mu}\tan\beta $ and Eq. (\ref{masses}) yields $M_2\sim(\tan^2\beta)2\times10^8$ GeV.
Provided $\tan\beta$ is large, it is then easier
to satisfy Eq. (\ref{conc}), especially for non-SUSY where there is no gravitino constraint on $T_r$.
For large $\tan\beta$ it also becomes possible to generate the BAU with the inflaton
decaying to $N_1$, as $M_1\sim(\tan^2\beta)5\times10^4$ GeV can satisfy Eq. (\ref{reh2}).

Thermal leptogenesis where the asymmetry is created by the decays of $N_2$ was
discussed in Refs. \onlinecite{DiBari:2005st,Engelhard:2006yg,Shindou:2007se} as well as
Refs. \onlinecite{Vives:2005ra,Hosteins:2006ja} which also relate $m_D$
to the up-type quark masses.
It is difficult to obtain sufficient asymmetry in this case.
For $\tilde{m}_{2,\tau}\approx m_{\rm atm}/2$, the bounds are
$T_r\gtrsim10^{10}$ GeV and $M_2\gtrsim5\times10^{10}$ GeV assuming that
$\epsilon_{2}$ is given by Eq. (\ref{davidson}) with $M_1$ replaced by $M_2$,
and that there is negligible washout from $N_1$
\cite{Buchmuller:2004nz,DiBari:2005st}. However, for quark-lepton symmetry
$\epsilon_2$ is suppressed by the lightest neutrino mass, 
and the phase values that maximize it do not coincide
with those that suppress the washout. We therefore expect these bounds to
be at least a few times larger. Similar conclusions are reached in Refs.
\onlinecite{Hosteins:2006ja}.
Notwithstanding the high $T_r$, the value of $M_2$ would then not be compatible with the
assumption $d_{D}\approx d_{u}$, although it may be compatible with 
$d_D\approx d_{\ell}\tan\beta$ for large $\tan\beta$. 

\section*{ACKNOWLEDGMENTS}
I thank Azar Mustafayev and Danny Marfatia for useful discussions.
I also thank L. Velasco-Sevilla and W. Rodejohann for helpful
comments and pointing out relevant papers. This research was supported by the U.S.
Department of Energy under Grant No. DE-FG02-04ER41308.

\end{document}